# Robust, scalable Hong-Ou-Mandel manifolds in quantum optical ring resonators

Edwin E. Hach, III[1], Stefan F. Preble[2], Ali W. Elshaari[3], Paul M. Alsing[4], and Michael L. Fanto[4]

[1]*School of Physics and Astronomy, Rochester Institute of Technology, Rochester, New York 14623, USA*
[2]*Microsystems Engineering, Rochester Institute of Technology, Rochester, New York 14623, USA*
[3]*Electrical and Electronic Engineering Department, University of Benghazi, Benghazi, Libya*
[4]*Air Force Research Lab, Rome New York 13441, USA*

## Abstract

Quantum Information Processing, from cryptography to computation, based upon linear quantum optical circuit elements relies heavily on the ability offered by the Hong-Ou-Mandel (H-O-M) Effect to "route" photons from separate input modes into one of two common output modes. Specifically, the H-O-M Effect accomplishes the path entanglement of two photons at a time such that no coincidences are observed in the output modes of a system exhibiting the effect. In this paper, we prove in principle that a significant increase in the robustness of the H-O-M Effect can be accomplished in a scalable, readily manufactured nanophotonic system comprised of two waveguides coupled, on-chip, to a ring resonator. We show that by operating such a device properly, one can conditionally "bunch" coincident input photons in a way that is far more robust and controllable than possible with an ordinary 50/50 beam splitter.

## I.     INTRODUCTION

The Hong-Ou-Mandel (H-O-M) Effect [1] is of obvious and ubiquitous importance to photonic quantum information processing [2]. For example, the design [3], implementation [4], and improvement [5] of the Knill, LaFlamme, Milburn (KLM)  Linear Optical Quantum Computing (LOQC) protocol inherently relies upon the conditional routing of coincident input photons that is the essence of the H-O-M effect.  Although there have been many impressive successful demonstrations of building blocks for the KLM protocol, such as those that have recently been realized in bulk optics [4],  there remains significant challenges towards realizing complex quantum information processing circuits and systems.  One key challenge that has recently been addressed is the need for miniaturization from bulk to integrated optics, [6], which allows for dense integration of circuit components and increasing scalability of quantum information systems. However, with the increasing level of integration of these devices it is crucial to have a process to overcome fabrication errors, whether this is thermal stabilization of the chip, a full chip calibration [7] or through the introduction of new devices, as we consider here, that can operate over wider operating conditions as well as offer denser integration and the ability to be reconfigured dynamically with low energy requirements.

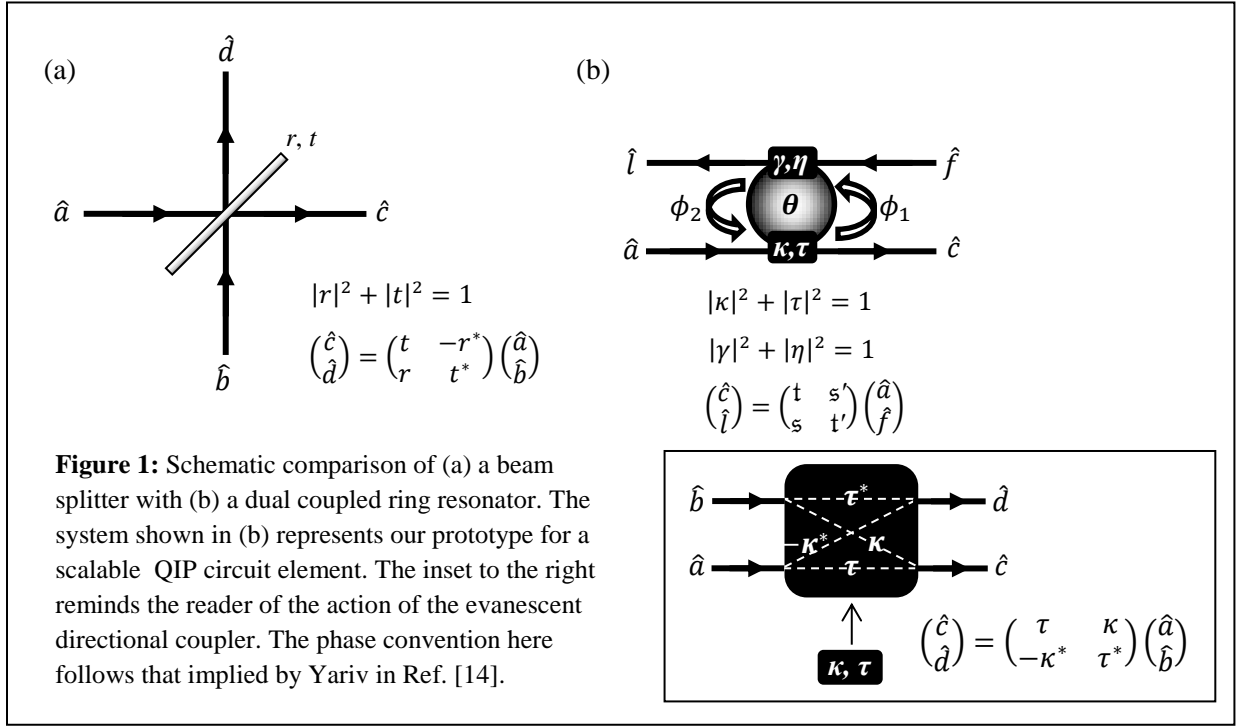

**Figure 1:** Schematic comparison of (a) a beam splitter with (b) a dual coupled ring resonator. The system shown in (b) represents our prototype for a scalable QIP circuit element. The inset to the right reminds the reader of the action of the evanescent directional coupler. The phase convention here follows that implied by Yariv in Ref. [14].

In this paper, we present a recent discovery, involving an on-chip, linear optical device, that, we believe, represents a substantial step toward scalable LOQC. Specifically, we show herein that a ring resonator system can, in principle, be operated efficiently to produce the H-O-M Effect over a far broader parameter range within a much richer parameter space than can systems based upon the more common architectures constructed from beam splitters and Mach-Zehnder Interferometers (MZI). Specifically, we show that the device shown in Figure (1b), which can be used as the fundamental building block of a scalable Nonlinear Sign (NS) shifter essential to the KLM protocol, exhibits, in principle, the H-O-M Effect, with 100% output state fidelity relative to the $|2::0\rangle$ NOON state, for infinite sets of parameter choices forming sub-manifolds of the parameter space of the device. In what follows, we shall refer to these sub-manifolds as Hong-Ou-Mandel Manifolds (HOMM). We emphasize that the existence of the HOMM within the device we consider is in sharp contrast to the situation in bulk optics where exact, 100% $|2::0\rangle$ NOON state fidelity exists at the output of a beam splitter on a hit-or-miss basis only for a single choice, viz. 50/50, of beam splitter parameters.

Ring resonators like the one shown schematically in Fig. (1b) are readily fabricated, and they have been shown to be of great importance for the realization of a wide range of integrated photonic devices and systems, from high performance electro-optic modulators to low-power densely integrated optical interconnects [8]. The relevant system parameters are the round trip linear phase shift, $\theta$, through the ring resonator, the cross-coupling parameters $(\kappa, \gamma)$, and the direct transmission parameters $(\tau, \eta)$ describing the two evanescent directional couplers, see the inset to Fig. (1). The couplers are assumed to be lossless, $|\kappa(\gamma)|^2 + |\tau(\eta)|^2 = 1$.

Single and two-photon transport analyses through systems like the one represented in Fig. (1b) have been presented in the literature. Notably, Shen and Fan have thoroughly examined the transport properties of such a system using a steady state scattering theory approach based upon Lipmann-Schwinger theory and the Bethe Ansatz [9, 10]. Also, some of the present authors have analyzed the quantum dynamics of single photon transport though such a system using an ansatz-free development along with Finite Difference Time Domain (FDTD) numerical integration [11]. In this letter, we describe the quantum optical system in the limit of continuous-wave (cw) operation. Working in this limit and using a discrete path integral approach [12,13] we can arrive at the quantum mechanical linear transformation relating the Boson operators along the output modes to those along the input modes,

$$\begin{pmatrix} \hat{c} \\ \hat{\imath} \end{pmatrix} = \begin{pmatrix} t & s' \\ s & t' \end{pmatrix} \begin{pmatrix} \hat{a} \\ \hat{f} \end{pmatrix} \Rightarrow \begin{pmatrix} \hat{a} \\ \hat{f} \end{pmatrix} = \begin{pmatrix} t^* & s^* \\ s'^* & t'^* \end{pmatrix} \begin{pmatrix} \hat{c} \\ \hat{\imath} \end{pmatrix} \Rightarrow \begin{pmatrix} \hat{a}^\dagger \\ \hat{f}^\dagger \end{pmatrix} = \begin{pmatrix} t & s \\ s' & t' \end{pmatrix} \begin{pmatrix} \hat{c}^\dagger \\ \hat{\imath}^\dagger \end{pmatrix}. \tag{1}$$

For convenience, we have defined the transition amplitudes

$$t \equiv \left(\frac{\eta^* - \tau e^{i\theta}}{\eta^*\tau^* - e^{i\theta}}\right), s \equiv \left(\frac{\gamma\kappa^* e^{i\phi_2}}{\eta^*\tau^* - e^{i\theta}}\right); t' \equiv \left(\frac{\tau^* - \eta e^{i\theta}}{\eta^*\tau^* - e^{i\theta}}\right), s' \equiv \left(\frac{\kappa\gamma^* e^{i\phi_1}}{\eta^*\tau^* - e^{i\theta}}\right) \tag{2}$$

using which it is straightforward to verify that Bosonic commutation relations are satisfied in the output mode operators. The form of this linear transformation is anticipated. Indeed, proceeding to the parametric limit, as we shall below, this result reduces to the corresponding classical result presented some years ago by Yariv [14].

While each belongs to the general class of U(2) linear optical devices, the fundamental advantage offered by a ring resonator over a beam splitter emanates from a difference in topology. To emphasize this point, we compare in Sec. II of this paper the operation of a lossless beam splitter, shown schematically in Fig. (1a) with that of the ring resonator device that is the subject of this letter, Fig. (1b). In the device represented in Fig. (1b), not only do the waveguide modes "scatter" as a result of local interactions with the directional couplers, but also they passively feedback on themselves at the couplers by virtue of the ring guide. This effect, referred to henceforth as Passive Quantum Optical Feedback (PQOF), is clearly not possible at the beam splitter in Fig. (1a). We include in

Sec. II a detailed discussion of PQOF, specifically relating its effects in the classical limit to a well known interference phenomenon arising from classical critical coupling between a waveguide and a lossy ring resonator.

The remainder of the paper is organized as follows. Within the framework of the theory of PQOF that we present in Sec. II, Sec. III is devoted to a systematic study of the existence and properties of the HOMM that are central subject of this work. Sec. IV is devoted to a discussion of the HOMM with respect to practical advantages they offer for applications in photonic Quantum Information Processing (QIP). We conclude the paper in Sec. V with remarks about the scope and possible applications of the results presented in the body of the paper. We indicate also in Sec. V some of the foci of our ongoing and future work on the quantum electrodynamics of ring resonator systems.

## II.     PASSIVE QUANTUM OPTICAL FEEDBACK (PQOF)

To illustrate the striking physical consequences that arise due to PQOF and to indicate the use of our formulation of the problem, we compare the devices shown in Fig, (1) assuming the input to be an ordinary coherent state injected along mode $a$. The input state in each case is $|\psi_{\text{in}}\rangle = \widehat{\mathcal{D}}^{(a)}(\alpha)|\emptyset\rangle$, where $\widehat{\mathcal{D}}^{(a)}(\alpha) = \exp(\alpha \hat{a}^\dagger - \alpha^* \hat{a})$ is the familiar Heisenberg-Weyl displacement operator [15], and $|\emptyset\rangle$ is the quantum electrodynamic vacuum.

In the case of the beam splitter, the output state is $|\psi_{\text{out}}^{\text{BS}}\rangle = \widehat{\mathcal{D}}^{(c)}(t\alpha)\widehat{\mathcal{D}}^{(d)}(r\alpha)|\emptyset\rangle$, where $r$ and $t$ are, respectively, the reflectance and transmittance of the beam splitter. It is clear from this form of the output state that, except in the trivial case in which $t = 0$, output mode $c$ will always be populated by a coherent field of non-zero amplitude; this is a standard result of elementary quantum optics [15].

Using Eqn. (1) it is straightforward to show that the output state in the case of the ring resonator is $|\psi_{\text{out}}^{\text{RR}}\rangle = \widehat{\mathcal{D}}^{(c)}(\mathfrak{t}\alpha)\widehat{\mathcal{D}}^{(l)}(\mathfrak{s}\alpha)|\emptyset\rangle$. In this case, the existence of PQOF allows for possibilities not available in the case of the beam splitter. In particular, the output field in mode $c$ can be suppressed even in non-trivial cases provided that $\mathfrak{t} = 0$. In fact, this result represents the quantum mechanical extension of the condition of critical coupling as has been predicted [14] and observed [16] in classically driven ring resonators.

In Fig. (2) we display schematically a lossy ring resonator driven classically via directional coupling with a single waveguide as in Fig. (2a) along with its appropriate quantum mechanical generalization, shown in Fig. (2b).

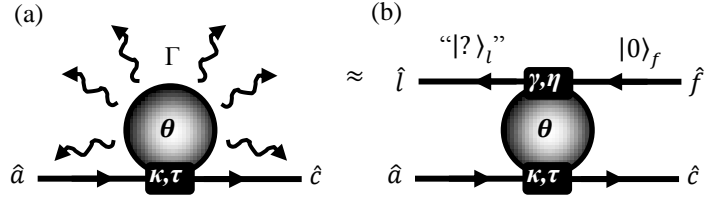

**Figure 2:** A cartoon-like representation of modeling a singly coupled lossy ring resonator (a) using a dually coupled, lossless ring resonator and drop port (b). Losses in (a) are modeled by tracing over mode $l$ in (b). This figure serves as a rationale for our choices of $f$ ("fictitious") and $l$ ("loss") for mode labels and operators in the system we analyze in this work.

Fig. (2b) is identical to Fig. (1b), except we emphasize in Fig. (2b) the role of the waveguide supporting modes $f$ and $l$ as a drop port for modeling losses in the ring resonator. To model, in a simple way, the losses from the ring resonator in Fig. (2a), we require that the input along the "fictitious" mode, $f$, in Fig. (2b) is necessarily the single mode vacuum. Incidentally, introduction of a drop port as we have done here is simply another example of the sort of unitary dilation necessary for the consistent treatment of linear quantum optical systems. In particular, a simple beam splitter, even when driven by a highly excited ordinary coherent state, must be equipped with a "fictitious" input port in order to preserve commutation relations [15]. In short, the quantum electrodynamic vacuum is not exactly nothing! Further, to model noise in the system, we must relinquish all knowledge about the quantum state of the output in the loss mode, $l$. Mathematically, we take an arbitrary input state along mode $a$ to the lossy ring resonator to be of the form $|\psi_{\text{in}}^{\text{LRR}}\rangle = \sum_m C_m (\hat{a}^\dagger)^m |\emptyset\rangle$, and we use Eqn. (1), effectively working in the Heisenberg picture [17], to determine the output state $|\psi_{\text{out}}^{\text{LRR}}\rangle = \sum_m C_m (t\hat{c}^\dagger + s\hat{l}^\dagger)^m |\emptyset\rangle$ from the dual waveguide system shown in Fig. (2b). We recover a quantum mechanical description of the output of the single waveguide system of Fig. (2a) along mode $c$ by first forming the global (pure state) density operator for the output of the dual waveguide system, $\hat{\varrho}_{\text{out}}^{(G)} = |\psi_{\text{out}}^{\text{LRR}}\rangle\langle\psi_{\text{out}}^{\text{LRR}}|$ and then tracing over the loss mode, $\hat{\varrho}_{\text{out}}^{(c)} = \text{Trace}_l\{\psi_{\text{out}}^{\text{LRR}}\rangle\langle\psi_{\text{out}}^{\text{LRR}}|\}$. Elsewhere, we shall apply this scheme of analysis in further detail to a wider range of systems of interest. We include the discussion here to provide a theoretical basis connecting our fully quantum mechanical analysis of the system shown in Fig. (2a) with previous results based on classical electrodynamics. In particular, we can now see that the loss rate from, or equivalently the cavity lifetime of, the ring resonator is related to the cross coupling parameter, $\gamma$, connecting the ring resonator to the drop port in Fig. (2b). Alternatively, we can follow Yariv in characterizing the same cavity loss information in terms of the intra-cavity "circulation" factor, which expresses that fraction of the classical amplitude that survives a round trip through the ring resonator [14]. Specifically, the parameter in our analysis corresponding to Yariv's circulation factor is $\eta^*$, which, for practical reasons we assume to be real, see below. For completeness, in Table (1) we present a precise transcription relating the parametric limit of our quantum mechanical analysis to the classical analysis presented in Ref. [14].

| Quantity | Yariv Classical Treatment Ref. [14] | Fully QM Treatment in Parametric Limit |
|---|---|---|
| Input Field Amplitude | 1 | $\alpha_{in}$ |
| Output Field Amplitude (waveguide) | $b_1$ | $\alpha_{out}$ |
| "CirculationFactor"/Internal Transmission Amplitude at Drop Port | $\alpha$ | $\eta^*$ |
| Round Trip Phase | $\theta$ | $-\theta$ |
| $a \rightarrow c$ transition Amplitude | $t$ | $\tau$ |

**Table 1:** Transcriptions between the classical variables used by Yariv in Ref. [14] and the parameters used in this paper in investigating the parametric limit of a driven ring resonator. Note that we have assumed $\eta \in \mathbb{R}$ for the present analysis.

To further expose the connection, consider the expectation value of the electric field in mode $c$. For the cw, traveling field mode we are considering, the field operator is $\hat{E}_x(z,t) = i\mathcal{E}_c\{\hat{c}e^{i(k_p z - \omega t)} - \hat{c}^\dagger e^{-i(k_p z - \omega t)}\}$, where we assume for notational simplicity and without loss of generality that the linearly ($x$-) polarized output field propagates in the $z$ direction. Considering the system in the parametric limit, we equate the quantum mechanical expectation value of the output field with the output of a classically driven system. This is allowed by virtue of the fact that we are working with factorizable products of quasi-classical ordinary coherent states; in other words, we may consider this a consequence of the optical equivalence theorem [15]. The output field is

$$E_x(z,t) = \langle \psi_{out}^{RR}|\hat{E}_x(z,t)|\psi_{out}^{RR}\rangle = i\mathcal{E}_c\{\mathfrak{t}\alpha e^{i(k_p z - \omega t)} - (\mathfrak{t}\alpha)^* e^{-i(k_p z - \omega t)}\} \tag{3}$$

It is clear from Eqns. (2) and (3) that the field in mode $c$ will vanish whenever $\mathfrak{t} \equiv \left(\frac{\eta^* - \tau e^{i\theta}}{\eta^* \tau^* - e^{i\theta}}\right) \rightarrow 0$. We note that in any case for which $|\eta| = |\tau| = 1$, the output field is merely phase shifted leaving the field amplitude in mode $c$ equal to that of the input field in mode $a$. Such a case is essentially trivial and corresponds to the physically uninteresting situation in which there is no cross-coupling from the input waveguide into the ring resonator.

Experimentally, the most important cases appear to be the ones in which both $\eta \in \mathbb{R}$ and $\tau \in \mathbb{R}$, so we proceed under these conditions, avoiding the trivial case mentioned in the previous paragraph. It is now clear that the effect of PQOF is encoded in the form of t. To see this, we examine the situation in which there is exact resonance between the feedback mode of the ring resonator and the input field, i.e. we set $\theta = 0 \mod 2\pi$. Under these conditions, we will show in the next paragraph that the output field in mode $c$ is suppressed as long as $\eta = \tau < 1$. These two conditions, exact resonance and balanced couplings, are precisely the conditions for critical coupling that lead to the destructive interference between that part of the field that is ***directly transmitted*** from mode $a$ to mode $c$ and the part that is ***fed back*** into mode $c$ after having been processed through the ring resonator. Classically, the destructive interference arises from the phase shift that occurs upon cross coupling (twice: once to get into the ring and again to get out) between the waveguide and the ring. The balanced coupling in the present case manifests classically as a "tuning" of the cavity losses in the ring so as to match the amplitudes of the direct and the feedback fields, assuring complete cancellation of the output field along mode $c$.

To understand the condition for destructive interference and to highlight the role of PQOF, consider the situation in Fig. (2a) from the perspective of classical electrodynamics. At exact resonance, the feed back field coupling out of the ring resonator and into mode $c$ will be exactly out of phase with the field directly transmitted through the directional coupler from mode $a$ to mode $c$. The reason for this is that the initial cross-coupling from mode $a$ into the ring resonator puts the cavity field in quadrature with the transmitted field, and then the second cross-coupling out of the ring resonator into mode $c$ puts the feedback field in opposition with the transmitted field. To demonstrate full cancellation of the fields; therefore, it remains to be shown that upon output, the feedback field and the directly transmitted fields have equal amplitudes. Assuming the input field amplitude along mode $a$ to be $\alpha_{\text{in}}$, ignoring the phases for the moment allowing us to write the cross coupling "factor" as $\sqrt{1-\tau^2}$, and replacing the effect of the drop port with the equivalent circulation factor, $\eta$, we proceed to compute the output amplitudes for the output fields interfering in mode $c$. For a field cross-coupling into or out of the ring $|\alpha| \xrightarrow{\text{cross}} \sqrt{1-\tau^2}\,|\alpha|$, for a field transmitted directly though the coupler, whether inside or outside of the ring, $|\alpha| \xrightarrow{\text{direct}} \tau\,|\alpha|$, and for a field making a single round trip inside the cavity $|\alpha| \xrightarrow{\text{r.t.}} \eta\,|\alpha|$. The output amplitude of the directly transmitted field is then $\alpha_{\text{out}}^{\text{dir}} = \tau \alpha_{\text{in}}$, where we assume for convenience that $\alpha_{\text{in}} \in \mathbb{R}$. The feedback field is a linear superposition of fields experiencing two cross-couplings and any number of round trips though the ring resonator, each of which picks up a factor of $\eta$ and all but the first of which picks up a factor of $\tau$ at the coupler,

$$\alpha_{out}^{fb} = \left(\sqrt{1-\tau^2}\right)^2 \{\eta\alpha_{in} + \eta\tau\eta\alpha_{in} + \eta\tau\eta\tau\eta\alpha_{in} + \cdots (\eta\tau)^j \eta\alpha_{in} + \cdots\} = (1-\tau^2)\eta\alpha_{in} \sum_{j=0}^{\infty}(\eta\tau)^j \quad (4a)$$

$$\alpha_{out}^{fb} = \left(\frac{1-\tau^2}{1-\eta\tau}\right)\eta\alpha_{in} \quad (4b)$$

Examining the ratio $\frac{\alpha_{out}^{fb}}{\alpha_{out}^{dir}} = \left(\frac{1-\tau^2}{1-\eta\tau}\right)\frac{\eta}{\tau}$ we see immediately that the output field amplitudes are equal for $\eta < 1$ and $\tau < 1$ if $\eta = \tau$, a direct enforcement of the critical coupling condition.

In this section we have introduced the concept of Passive Quantum Optical Feedback (PQOF) as manifested in the ring resonator system of Figs. (1b) and (2b) and as described mathematically via Eqns. (1) and (2). By way of a simple comparison with a standard beam splitter, we have demonstrated a topological advantage due to PQOF for the ring resonator system in comparison with other U(2) linear optical devices which are not capable of PQOF. Specifically, we have demonstrated that, in the classical limit, it is PQOF that is ultimately responsible for complete destructive interference in a ring resonator driven in the condition of critical coupling via a waveguide. We now turn to a decidedly more quantum mechanical feature of PQOF that we anticipate will have important and far reaching applications in schemes for photonic QIP.

## III.   HONG-OU-MANDEL MANIFOLDS (HOMM)

Using Eqns. (1) and (2), it is straightforward to establish the existence of the HOMM in our system. Referring again to Fig. (1b), we consider the usual input for the H-O-M Effect along modes $\hat{a}$ and $\hat{f}$, $|\psi_{in}\rangle = \hat{a}^\dagger\hat{f}^\dagger|\emptyset\rangle = |1,1\rangle$. In the output modes we find $|\psi_{out}\rangle = \sqrt{2}\mathfrak{ss}'|2,0\rangle + (\mathfrak{ss}' + \mathfrak{tt}')|1,1\rangle + \sqrt{2}\mathfrak{t}'\mathfrak{s}|0,2\rangle$. It is clear upon inspection of this result that **the H-O-M Effect obtains as long as the constraint $\mathfrak{ss}' + \mathfrak{tt}' = 0$ is satisfied**; under this condition the probablilty $P(1,1) = |\mathfrak{ss}' + \mathfrak{tt}'|^2$, for photon coincidence in the output modes, $c$ and $l$, vanishes. We shall refer to this constraint as the strong form of the Hong-Ou-Mandel Manifold Constraint (sHOMMC); the existence and important features of these manifolds are the central results presented in this paper.

To appreciate the importance of the HOMM, consider the parameter space of the device pictured in Fig. (1b) in comparison with that of an ordinary, lossless beam splitter as in Fig. (1a). In the former case, the four complex parameters $\kappa$, $\tau$, $\gamma$, and $\eta$, along with the single real parameter $\theta$ characterize a nine dimensional manifold that is the parameter space of the system. Of these nine dimensions, the four phases of the coupling parameters are essentially redundant. In practice, these phases are typically determined by the details of the construction of the coupler and thusly do not represent truly "free" parameters within the operating space of the device. In any case, these phases can be adjusted either on-chip or in bulk optics via linear phase shifters. For analogous reasons, the general operation of a beam splitter can be described without loss of generality using a one dimensional parameter space, the H-O-M Effect occurring for an individual (50/50) operating point (i.e. a zero dimensional "HOMM") within that one dimensional parameter space. As we now examine, the sHOMMC establishes the existence of higher (>0) dimensional manifolds that are level sets of operating points for which the device shown in Fig. (1b) exhibits the H-O-M Effect.

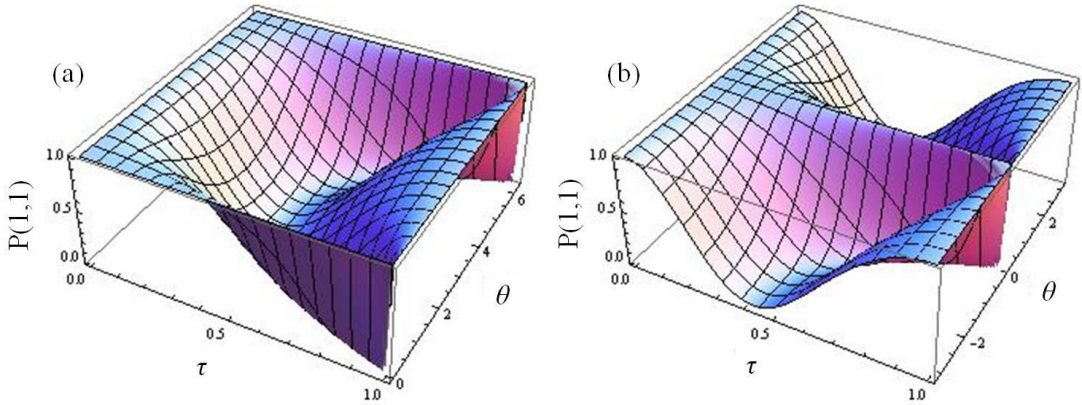

**Figure 3:** Surface plot of the probability, $P(1,1)$, for output photon coincidence for the case in which two incident photons are coincident on the ring along modes $a$ and $f$. The Hong-Ou-Mandel Effect are the zeroes of this surface plot. The level set of zeroes forms a continuous manifold in this case; it is a one dimensional Hong-Ou-Mandel Manifold (HOMM). In this figure, as in Figs. (4) and (5), we display the same result over the ranges (a) $\theta \in [0, 2\pi]$ and (b) $\theta \in [-\pi, \pi]$. We do this merely for ease of visualization and reference; the two plots in each figure so displayed convey exactly the same information.

To clarify the result we examine case in which the device shown in Fig. (1b) is characterized via $\tau = \eta \in \mathbb{R}$ and $\kappa = \gamma = ik, k = \sqrt{1 - \tau^2} \in \mathbb{R}$. We note that these constraints, while restrictive, are experimentally relevant, and devices adhering to them have been fabricated by our group [18] and by others [19]. This constraint corresponds at exact resonance with the optical balance condition for critical coupling. In this case, the parameter space of the device is restricted to two dimensions. Here we take the free parameters to be $\tau$ and $\theta$. Applying these constraints to the sHOMMC and using Eqn. (2), we plot in Figure (3) the probability of output photon coincidence in modes $l$ and $c$ versus the free parameters of the system. It is precisely the bottom of the "valley of zeroes" of the $P(1,1)$ that is the one dimensional HOMM for this system. In obvious contrast to the 50/50 beam splitter, which has a single H-O-M operating point, the system shown in Fig. (1b) has, in this case, infinitely many such operating points. This level set of vanishing probability for photon coincidence is the one dimensional HOMM within the two dimensional parameter space of the system in this case.

To further investigate the one dimensional HOMM for the system we consider a weaker form of the sHOMMC, which we will call wHOMMC, written in terms of the system parameters as

$$|\kappa|^2 + |\gamma|^2 + |\kappa|^2|\gamma|^2 + 2\text{Re}(\eta\tau e^{i\theta}) = 2 \tag{5}$$

Eqn. (5) is obtained from the sHOMMC by using Eqn. (2) and requiring the **numerator** of the expression to vanish. Working within the foregoing special case, we can solve Eqn. (5) in closed form to obtain analytically a parametric curve for the HOMM. The physical branch of the solution can be written as

$$\tau(\theta) = \sqrt{2 - \cos(\theta) - \sqrt{(2 - \cos(\theta))^2 - 1}} \tag{6}$$

We plot Eqn. (6) in Figure (4).

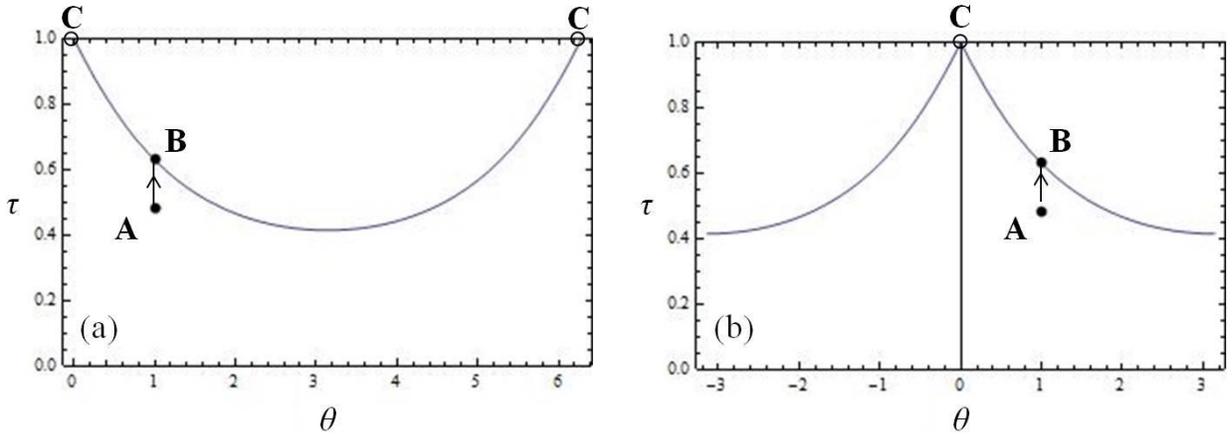

**Figure 4:** Parametric plot of a one dimensional HOMM based upon the wHOMMC. The point(s) labeled C is the one for which the wHOMMC breaks down. Away from Point(s) C, the parametric curve shown in this figure lies along the level set of zeroes in Fig. (3). Points A and B are included to accompany the discussion in Sec. IV. In (a) $\theta \in [0, 2\pi]$ and in (b) $\theta \in [-\pi, \pi]$.

Note that Fig. (4) correctly represents the HOMM for the system except in the limit $\theta \to 0 \mod 2\pi$, $\tau \to 0$, that is, the points labeled C in that figure must be disregarded. One can show that these are the only points within the parameter space of the system for which the **denominator** of the sHOMMC vanishes, thusly forcing us to take the limit using the L'Hôpital rule; the result is $P(1,1) \to 1$, as expected for perfect transmission through each of the directional couplers and as can be seen upon inspection of Fig. (3). Except for at the endpoints, the curve shown in Fig. (4) can be inlaid exactly along the bottom of the $P(1,1) = 0$ level set of the surface shown in Fig. (3).

It is noteworthy to mention that in this simple case, perfect photon coincidence is expected whenever the field is exactly resonant with a mode of the ring. That is, whenever $\frac{nR\omega}{c} \epsilon \mathbb{N}$, where $R$ is the radius and $n$ the linear index of refraction of the resonator and $\omega$ is the angular frequency of the field, such that $\theta = 0 \mod 2\pi$, it is straightforward to show that $P(1,1; \tau) = 1 \ \forall \ \tau \in [0,1]$. This behavior is also evident in Fig. (3). Appealing to the semi-classical argument presented in the previous section, we understand this limiting behavior at resonance as follows. Feynman paths connecting either input mode to the corresponding output mode along the same waveguide interfere destructively; recall that we are currently imposing conditions of critical coupling. Now, however, we must include contributions from Feynman paths that describe exchange between waveguides, and the amplitudes describing the various exchange paths (j + "½" round trips in the ring resonator – quotes to remind the reader that the result is insensitive to the exact placement around the ring of either of the directional couplers relative to the other) do not interfere destructively, allowing for the observed photon coincidences.

Suppose that we now relax the restriction that the directional couplers are balanced; we allow $\tau \neq \eta$ though we continue to require both to be real. Under these more relaxed constraints, the system is now characterized by a three dimensional parameter space in which we take the free parameters to be $\eta, \tau$ and $\theta$. In Figure (5) we plot, using the sHOMMC, the surface on which the probability of photon coincidence in the output modes $l$ and $c$ vanishes. The surface shown in Fig. (5) is the two dimensional HOMM for this slightly more complex situation. Cutting the plot shown in Fig. (5) with the appropriate diagonal plane (i.e. the one on which $\eta = \tau$) through the cubic frame of the figure reproduces the curve for physical root for $\tau = \eta$ vs. $\theta$ shown in Fig. (4), again neglecting the pathological behavior of the endpoints of the curve in Fig. (4).

It is clear from the examples presented in this section that the device shown in Fig. (1b) admits a rich subset of its parameter space on which the H-O-M Effect occurs. The structure of each HOMM is directly related to the form of the quantum mechanical mode transformation given in Eqns. (1) and (2). The form of this transformation, in turn, encodes the PQOF inherent to the topology of the system. Without PQOF, as in the case of the beam splitter shown in Fig. (1a), the HOMM is a single point. With PQOF, as in Fig. (1b), there are infinitely many H-O-M operating points that form continua within the parameter space of the device. Constructing a simple device for which the dimensionality of the HOMM is non-zero seems to rely upon the existence or lack of PQOF amongst the various Feynman paths through the device. We anticipate that this trend will make for even more interesting quantum interference phenomena as we extend, elsewhere, our analysis to U(N) linear optical devices for N > 2.

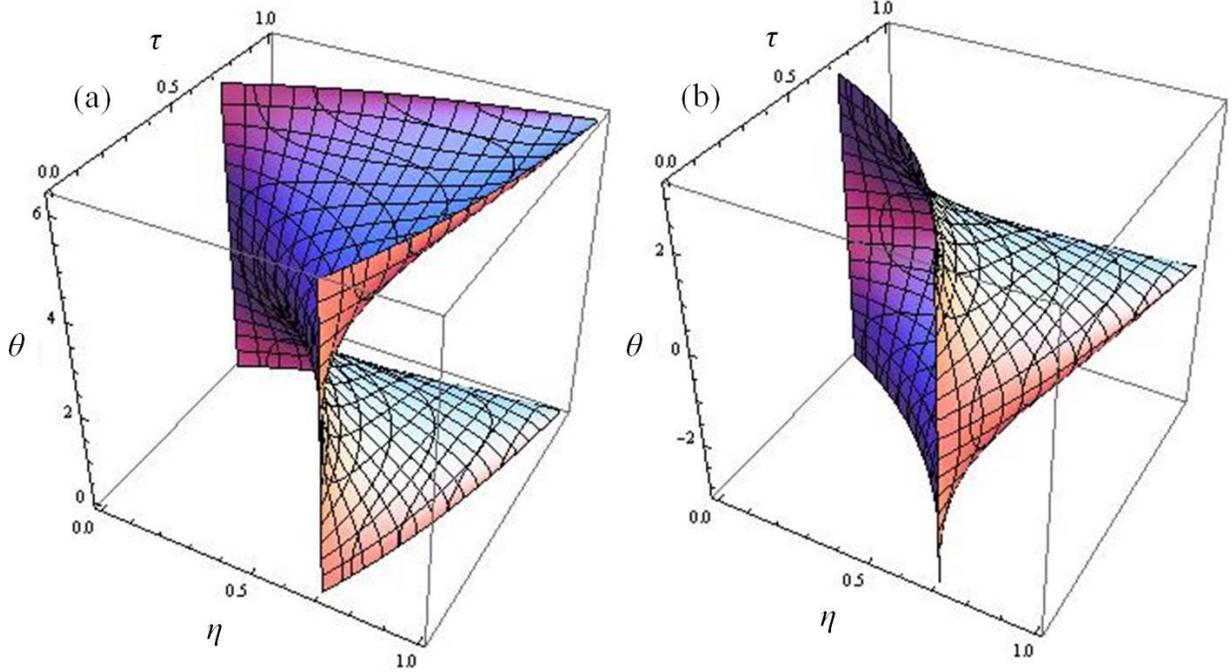

**Figure 5:** Surface plot of the two dimensional HOMM for the case of an unbalanced dually coupled ring resonator. In (a) $\theta \in [0,2\pi]$ and in (b) $\theta \in [-\pi,\pi]$. The higher dimension of the HOMM for this case suggests even more opportunity for robustness of the H-O-M along the lines of that described for the one dimensional case discussed explicitly in the paper.

Even if just as a simple matter of a priori statistics, a device having more successful operating points within its parameter space is expected to more robustly succeed in producing the desired effect via its operation. However, the situation, from a device perspective, is even better than that. In fact, the inherent scalability of devices such as the one shown in Fig. (1b) along with the tunability of couplings within the device can be used in conjunction with the HOMM that we have demonstrated in this section to produce a dynamic robustness which, when integrated on-chip, could form an integral part of a new generation of QIP circuits. We proceed in the next section to discuss in more detail some of these possibilities.

## IV. RING RESONATORS AS QIP CIRCUIT ELEMENTS

The broader importance of the results we have presented here is really two-fold. First, we have demonstrated that a ring resonator evanescently coupled to two waveguides as shown in Fig. (1b) is a system capable of "routing" input photons in accordance with the well-known Hong-Ou-Mandel Effect. Unlike the situation in bulk optics; however, the system we have analyzed here is scalable [20] and can be fabricated in large numbers [21]. These attributes suggest that such system could be easily engineered for on-chip implementation of the H-O-M Effect, a staple of many, if not all, quantum information processing architectures based upon linear quantum optics. In fact, one can readily establish that the system shown in Fig. (1b) is formally equivalent to an (scalable, easily manufactured) externally driven Faby-Perot etalon [22, 23].

Second, we have demonstrated that the range of parameters for which the ring resonator system demonstrates H-O-M behavior is infinitely larger than the corresponding range in the typical beam splitter version of the effect. This fact, combined with the dynamical tunability of the waveguide/ring resonator coupling [24] allows for the possibility of robust photon "routing" unlike that possible with a beam splitter. To see this, suppose that a balanced device like the one shown in Fig. (1b) is operating within its two dimensional parameter space at Point A in Fig. (4). Clearly this point is not within the HOMM for the system, and we should expect to observe photon coincidences in the output modes. This situation corresponds to the behavior one observes in an unbalanced beam splitter. However, unlike the case of the beam splitter we can dynamically "correct' the situation by tuning, thermally for example, the coupling between the waveguides and the ring. Such a "correction" corresponds in Fig. (4) to a vertical translation of the operating point of the system, ideally to Point B shown explicitly on the HOMM curve. In this sense, the H-O-M Effect is more robust in the ring resonator set-up than it is in its beam splitter implementation. What we have demonstrated here is that, by virtue of the existence of the HOMM within the parameter space of the waveguide/ring resonator system, ***we can produce on-chip, scalable, robust quantum optical interconnects for integration into photonic QIP devices***.

We conclude this section with a brief discussion of the operationally advantageous design parameters for such devices; specifically we wish to start the process of homing in on the "sweet" region for practical operation of the device we have analyzed above. To motivate the discussion, we plot in Fig. (6) $P(1,1)$ versus $\theta$ for several values of $\tau$. We assume again that $\eta = \tau \in \mathbb{R}$ and, for ease of reference, we consider the plot in Fig. (6) over the interval $\theta \epsilon [-\pi, \pi]$; each plot in Fig. (6) is a vertical cut taken from Fig. (3b).

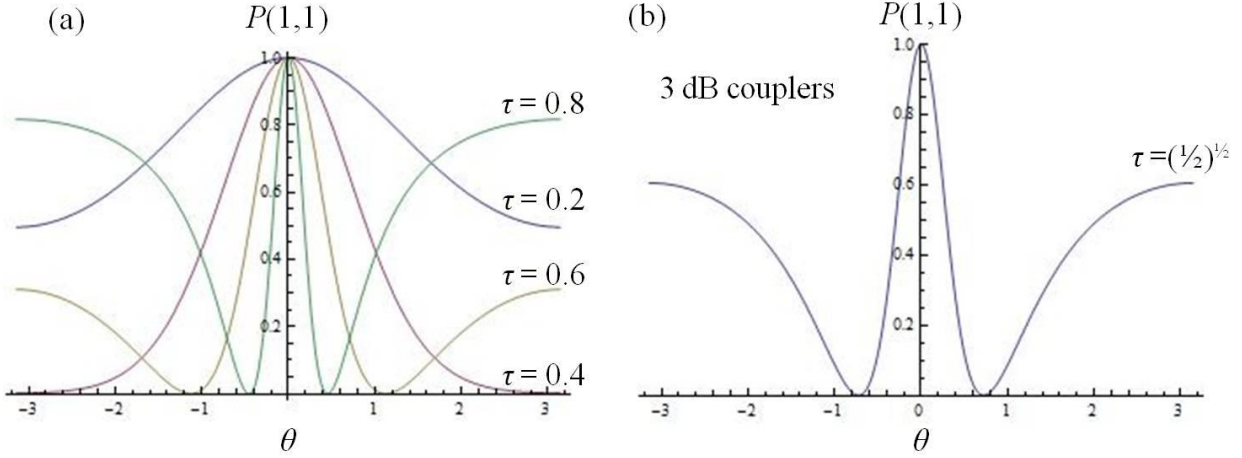

**Figure 6:** Plots of $P(1,1)$ for several values of $\tau = \eta \in \mathbb{R}$, corresponding to balanced operation of the dually driven ring resonator. The H-O-M dips are apparent, provided they occur. Clearly, both the existence and resolution of such and H-O-M dip are dependent upon the coupling parameters for the system. This suggests the existence of an "optimal" operating region for an on-chip device made from dually coupled rings. In (b) we demonstrate that the pragmatically important case involving 3dB couplers is safely within this "optimal" region, however heuristically defined at this point.

Under conditions in which the effect occurs, the H-O-M dips in Fig. (6) are symmetrically placed about $\theta = 0$. Two important observations are clear. First, there is a cutoff value for $\tau$ beneath which the H-O-M Effect does not occur in the operation of the device for any relationship between the field frequency and the frequency of the mode of the ring resonator. Using Eqn. (5) it is easy to show for this case that the cutoff value is $\tau_{\text{cutoff}} = \sqrt{3 - \sqrt{8}} \approx 0.4142$, in agreement with Fig. (6a). Practical concerns, as discussed in Section III, above, restrict the operation of the device in the regime $\tau \lesssim 1$. For $\tau = 1 - \varepsilon$, the H-O-M dips become arbitrarily narrow and close to $\theta = 0$ for arbitrarily small but non-zero values of $\varepsilon$. For $\varepsilon = 0$, the H-O-M dips suddenly disappear (see Sec. III), making the regime near $\tau = 1$ impracticable for device operation and mathematically trivial. Referring to Fig. (6a), it is clear that values of $\tau$ equal to or even in excess of 0.8 are quite practical for operation; the H-O-M dips are obviously visible and resolvable. Based upon these observations, we take the outer limits on the practical operating regime for the tunable H-O-M Effect in the device in the balanced case we consider here to be $\sqrt{3 - \sqrt{8}} < \tau \lesssim 0.8$. Of course, upper and lower limits of this regime are chosen rather heuristically, and depending upon the details of a particular application, we expect to find more rigid empirical foundations for and corrections to them. We defer detailed analysis of special cases to forthcoming papers specifically about those applications. Instead, we point out that the experimentally important case of the balanced ring resonator with 3dB couplers, $\tau = \eta = \frac{1}{\sqrt{2}}$, is safely within the practically attainable H-O-M regime for the device; the H-O-M dips for this important special case are clearly resolvable in Fig. (6b).

## V. SUMMARY AND OUTLOOK

In this paper, we have developed the theory of Passive Quantum Optical Feedback (PQOF), and we have presented examples of important physical consequences of it. In particular, we have predicted the existence of Hong-Ou-Mandel Manifolds (HOMM) within the parameter space of a PQOF Device (PQOFD), namely, the externally driven ring resonator system shown schematically in Fig. (1b).

In the interest of conservative reporting of our findings, we remark in closing that the comparison between a beam splitter and the PQOFD discussed above, while fair on a practical device level, must be qualified in the following sense. The beam splitter, Fig. (1a), just as the individual directional coupler (see inset to Fig. (1)) induces mode transformations in the coset space SU(2)/U(1). The PQOFD induces mode transformations in the slightly less restricted coset space U(2)/U(1). It may, therefore, be tempting to dismiss, prematurely as it turns out, the PQOFD as merely another type of Mach-Zehnder Interferometer (MZI). The situation is not quite that simple. Again, owing to PQOF, the PQOFD topologically "unpacks" the system relative to the MZI such that quantum interference effects (e.g. the H-O-M Effect) can be made to occur via global tuning of the system (viz. tuning the balanced couplers of the PQOFD) as opposed merely the differential tuning (viz. phase shifting one arm) available in the single pass MZI. This attribute only enhances the possibility for miniaturizing and integrating the PQOFD for on-chip applications.

Our ongoing work is partially focused on designing and demonstrating specific devices for applications in QIP. One specific example essential to the KLM protocol for LOQC is that of the NS gate. We have discovered a scalable design, and we shall report this result elsewhere very soon.

**Acknowledgements**

Support for this work was provided by the Air Force Research Lab (AFRL) Visiting Faculty Research Program (VFRP) Grant SUNY-IT #FA8750-13-2-0115. Some of the authors were also supported by Dr.